\RequirePackage{fix-cm}
\documentclass{llncs}
\usepackage{graphicx}
\usepackage{color}
\usepackage{soul}
\usepackage{tabularx}
\usepackage{booktabs}
\usepackage[exponent-product=\cdot]{siunitx}
\usepackage{hyperref}
\usepackage{pgf}
\usepackage{tikz}
\usepackage{multirow}
\usepackage{subcaption}
\begin{document}

\title{
Uncertainty-aware performance assessment \\
of optical imaging modalities \\
with invertible neural networks
}


\author{Tim J. Adler\(^1\)         \and
        Lynton Ardizzone\(^2\) \and
        Anant Vemuri\(^1\)\and
        Leonardo Ayala\(^1\)\and
        Janek Gr\"{o}hl\(^1\)\and
        Thomas Kirchner\(^1\)\and
        Sebastian Wirkert\(^1\)\and
        Jakob Kruse\(^2\)\and
        Carsten Rother\(^2\)\and
        Ullrich K\"{o}the\(^2\)\and
        Lena Maier-Hein\(^1\)
      } 
      
 

\institute{{1)} Computer Assisted Medical Interventions, German Cancer Research Center, Heidelberg, Germany\\
  {2)} Heidelberg Collaboratory for Image Processing, Heidelberg University,
  Heidelberg, Germany
}


\maketitle

\begin{abstract}
\textit{Purpose}: Optical imaging is evolving as a key technique for advanced sensing in the operating room. Recent research has shown that machine learning algorithms can be used to address the inverse problem of converting pixel-wise multispectral reflectance measurements to underlying tissue parameters, such as oxygenation. Assessment of the specific hardware used in conjunction with such algorithms, however, has not properly addressed the possibility that the problem may be ill-posed.\\
\textit{Methods}: We present a novel approach to the assessment of optical imaging modalities, which is sensitive to the different types of uncertainties that may occur when inferring tissue parameters. Based on the concept of invertible neural networks, our framework goes beyond point estimates and maps each multispectral measurement to a full posterior probability distribution which is capable of representing ambiguity in the solution via multiple modes. Performance metrics for a hardware setup can then be computed from the characteristics of the posteriors. \\
\textit{Results}: Application of the assessment framework to the specific use case of camera selection for physiological parameter estimation yields the following insights: (1) Estimation of tissue oxygenation from multispectral images is a well-posed problem, while (2) blood volume fraction may not be recovered without ambiguity. (3) In general, ambiguity may be reduced by increasing the number of spectral bands in the camera. \\ 
\textit{Conclusion}: Our method could help to optimize optical camera design in an application-specific manner.

\end{abstract}


\section{Introduction}
\label{intro}

\begin{figure}[htbp]
    \centering
    \begin{subfigure}[b]{0.48\textwidth}
        \centering
        \includegraphics[scale=0.232]{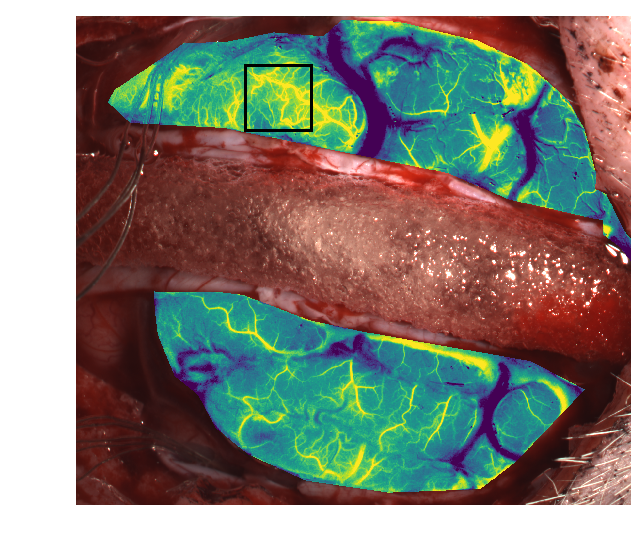}
       \caption{ }
       \label{fig:pig-oxygenation}
    \end{subfigure}
    \begin{subfigure}[b]{0.48\textwidth}
        \centering
        \includegraphics[width=\textwidth]{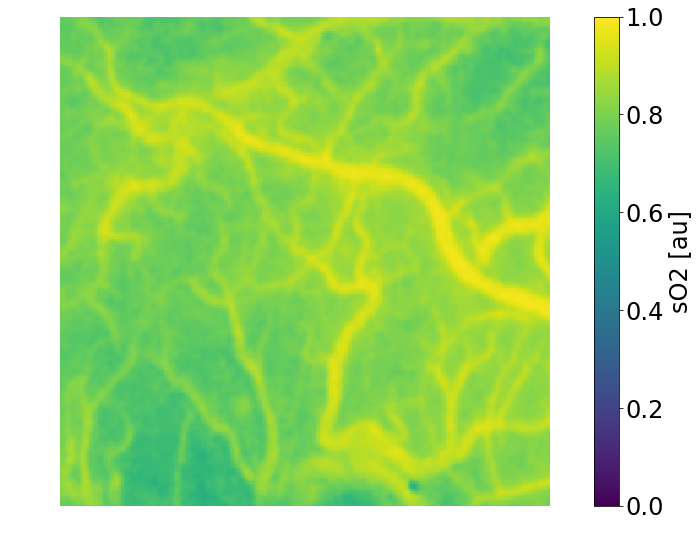}
       \caption{ }
       \label{fig:pig-oxygenation-patch}
    \end{subfigure}
    \caption{Oxygenation map of a porcine brain generated by 8-band multispectral imaging. (a) The full brain. (b) Cropping of the region marked in (a).}
    \label{fig:oxygenation-map}
\end{figure}

Many key challenges in the intersection of natural sciences and the life sciences are related to solving \emph{inverse problems}. Here, it is assumed that a forward process maps the (hidden) parameters of interest $\mathbf{x} \in X$ to observations $\mathbf{y}\in Y$ that can be measured. In the context of computer assisted interventions (CAI), for example, $\mathbf{x}$ may refer to important physiological tissue parameters, such as the tissue oxygenation $sO_2$ (cf.\ Figure~\ref{fig:oxygenation-map}), while $\mathbf{y}$ may represent multispectral measurements of tissue. The problem is usually solved by regression, which gives a point estimate for the tissue parameter(s) of interest based on the camera measurements~\cite{clancy_intraoperative_2015,wirkert_physiological_2017,wirkert_robust_2016}. However, in most inverse problems the mapping between $X$ and $Y$ is not injective, and two substantially different $\mathbf{x_1},\mathbf{x_2} \in X$ can result in the same $\mathbf{y}$. 
To recover a unique inverse, a regularizer can be added to the objective, but this approach, although commonly used, neglects the inherent ambiguity of the solution.
For our application, an explicit analysis of the ambiguity is crucial to identify the most suitable camera in terms of number and characteristics of camera bands. 
To our knowledge, none of the existing parameter estimation methods has incorporated a sufficiently powerful uncertainty quantification to do so.  

\begin{figure}[htbp]
    \centering
    \begin{subfigure}[b]{0.48\textwidth}
        \centering
        \includegraphics[width=\textwidth]{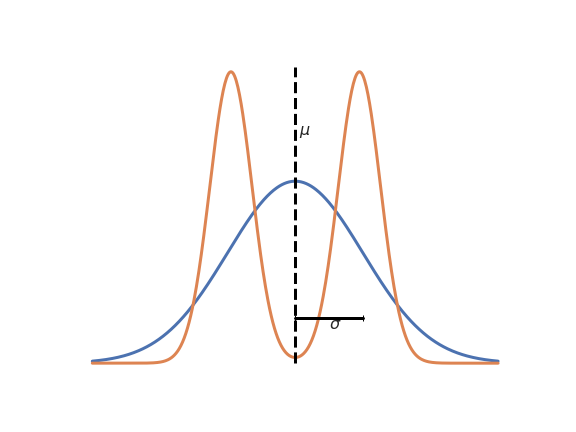}
       \caption{ }
       \label{fig:explain:multimode_vs_singlemode}
    \end{subfigure}
    \begin{subfigure}[b]{0.48\textwidth}
        \centering
        \includegraphics[width=\textwidth]{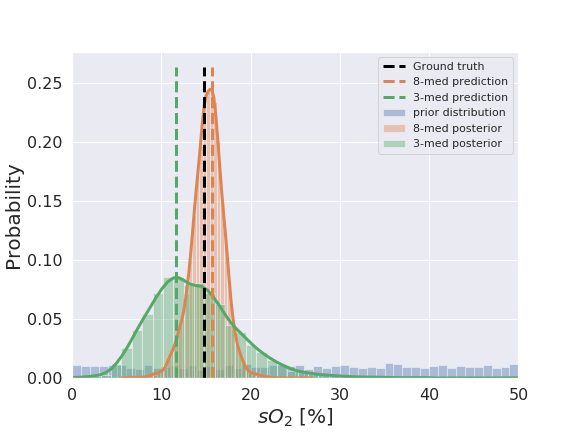}
        \caption{ }
        \label{fig:histogram}
    \end{subfigure}
    \caption{(a) Example of a unimodal (blue) and bimodal (orange) distribution with the same expectation value $\mu$ and variance $\sigma^2$. (b) Example of two posterior distributions as provided by our INN. The  posterior of the 3-band camera (green) is multi-modal, and the MAP estimation of tissue oxygenation is associated with the wrong mode leading to a poor estimation. The posterior of the 8-band camera (orange) is uni-modal with small width of the mode and better MAP estimation.}
\end{figure}

Current approaches to uncertainty quantification in the field of deep learning, such as dropout sampling (cf. e.g. \cite{gal_dropout_2016,leibig_leveraging_2017,li_dropout_2017,srivastava_dropout_2014}), probabilistic inference (cf. e.g. \cite{feindt_neural_2004,kohl_probabilistic_2018,zhu_bayesian_2018}) or ensembles of estimators (cf. e.g. \cite{lakshminarayanan_simple_2017,smith_understanding_2018}), typically augment traditional point estimates with confidence intervals, but do not recover unrestricted full posteriors $p(\mathbf{x} | \mathbf{y})$.
Consequently, these methods do not account for the possibility that the same observation $\mathbf{y}$ corresponds to {\em fundamentally different} $\mathbf{x}$. In other words these methods would always assume that $\mathbf{x}$ follows the blue (unimodal) distribution depicted in Figure~\ref{fig:explain:multimode_vs_singlemode} even if $\mathbf{x}$ followed the orange (multimodal) distribution.
The following two cases illustrate that this is a serious shortcoming when we wish to recover a physiological parameter $\mathbf{x}$ from observations $\mathbf{y}$ (Figure~\ref{fig:histogram}):
\begin{enumerate}
\item The solution is unique but suffers from high uncertainty. This may be represented by a uni-modal posterior $p(\mathbf{x} | \mathbf{y})$ whose single mode has large standard deviation. 
\item The problem is ill-posed in the sense that two substantially different $\mathbf{x_1},\mathbf{x_2}$ yield the same $\mathbf{y} = f(\mathbf{x_1}) = f(\mathbf{x_2})$. This must be represented by a multi-modal posterior $p(\mathbf{x} | \mathbf{y})$ whose individual modes may have low uncertainty. 
\end{enumerate}

%
Forcing a uni-modal representation onto the second case cannot work: It would either focus on one of the modes and miss the other, or cover both solutions under a single wide mode (similar to case 1) whose maximum is located at the average of $\mathbf{x_1}$ and $\mathbf{x_2}$ -- a highly implausible $\mathbf{x}$-value for the given $\mathbf{y}$.


We therefore argue that an ideal method for comparative camera assessment should be able to deal with \textit{all} possible types of uncertainty. 
We propose to move beyond point estimates by mapping an observation $\mathbf{y}$ to a full distribution $p(\mathbf{x} \,|\, \mathbf{y})$ rather than a single $\mathbf{x}$. 
To this end, we solve the resulting inverse problem using the recently proposed concept of \textit{invertible} neural networks (INNs)~\cite{ardizzone_analyzing_2018}. 
Performance measures for a hardware setup can then be computed from the number and widths of the modes of the posteriors, as illustrated in Figure~\ref{fig:histogram}. 

In the following sections, we describe our approach in detail and apply it to the comparative assessment of four different camera designs given the specific use case of physiological parameter estimation from multispectral imaging data. 

\section{Methods}
\label{methods}
In this section, we formalize the proposed approach to performance assessment in a generic manner and apply it to the specific use case of camera selection for multispectral image analysis.

Generally speaking, we assume that the method to be assessed involves a hardware setup $H$ (e.\,g.\ a multispectral camera) that is used to solve an inverse problem with a well-known forward process $f: X \to Y$, such as the mapping of tissue oxygenation $\mathbf{x}$ to the pixel-wise measurement $\mathbf{y} = f(\mathbf{x})$ of a multispectral camera.  We further assume that we have access to a data set $T = T^\text{train} \cup T^\text{validation} \cup T^\text{test} $ composed of tuples $(\mathbf{x},\mathbf{y})$, with $\mathbf{y} = f(\mathbf{x})$. Typically $T$ can be generated by means of Monte Carlo simulation, as in~\cite{kirchner_context_2018,wirkert_robust_2016,wirkert_physiological_2017} assuming the (virtual) hardware setup $H$. Finally, we represent the regressor $r$ as an invertible neural network, as detailed in Section \ref{ssec:INN}.

Our approach to performance assessment involves the following steps: (1) Training the regressor $r$ on $T^\text{train}$ using $T^\text{validation}$ for hyperparameter tuning. (2) Applying $r$ to $T^\text{test}$, to get a target distribution $p(\mathbf{x} \,|\, \mathbf{y})$ for each $\mathbf{y}$ in the test data set. (3) Extracting the modes for each $p(\mathbf{x} \,|\, \mathbf{y})$. (4) Computing descriptive statistics over the number and widths of the modes to quantify the uncertainty of the regressor. 
Different hardware setups can then be compared using metrics that consider not only the accuracy but also the uncertainty characteristics of the regressor. The following paragraphs instantiate this approach in the specific context of camera selection for intra-operative physiological parameter estimation.

\vspace{-0.5cm}
\subsection{Data generation for performance assessment} 
\label{ssec:data}

 We apply Monte Carlo methods to generate tuples of physiological parameters $\mathbf{x}$ and corresponding pixel-wise measurements $\mathbf{y} = f(\mathbf{x})$. The method is based on previous work \cite{wirkert_physiological_2017} and briefly revisited here. 
 
 Tissue is assumed to be composed of three infinitely wide layers. Each layer is defined by the following tissue parameters: blood volume fraction $v_{\text{HB}}$, reduced scattering coefficient at 500nm $a_{\text{mie}}$, scattering power $b_\text{mie}$, anisotropy $g$, refractive index $n$ and layer thickness $d$. Based on values for hemoglobin extinction coefficients $\epsilon_{\text{HB}}$ and $\epsilon_{\text{HBO2}}$ from literature \cite{jacques_optical_2013}, absorption and scattering coefficients $\mu_a$ and $\mu_s$ have been determined for use in the MC simulation framework. A Graphics Processing Unit (GPU) accelerated version \cite{alerstam2010next} of the Monte Carlo Multi-Layered (MCML) simulation framework \cite{wang1995mcml} was chosen to generate spectral reflectances. The spectral reflectances $r(\lambda)$ as determined by the MC simulation can be transformed to the reflectance measurement $r_i$ at band $i$ of a given camera $c$ by:
\begin{equation}
r_i=\frac{\int_{\lambda_{\text{min}}}^{\lambda_{\text{max}}}o(\lambda)l(\lambda)f_i(\lambda)r(\lambda)\,\text{d} \lambda }
{\int_{\lambda_{\text{min}}}^{\lambda_{\text{max}}}o(\lambda)l(\lambda)f_i(\lambda)\,\text{d} \lambda }
\label{eq:camera_reflectance}
\end{equation}
Here, the camera $c$ is characterized by $f_i(\lambda)$, the $i$th filter response, $l(\lambda)$, the relative irradiance of the light source and $o(\lambda)$, which represents other parameters of the optical system like camera quantum efficiency or transmission of the optical elements. 

\vspace{-0.5cm}
\subsection{Invertible Neural Networks (INNs) for physiological parameter estimation}
\label{ssec:INN}


\begin{figure}
    \centering
    \includegraphics[width=0.8\textwidth]{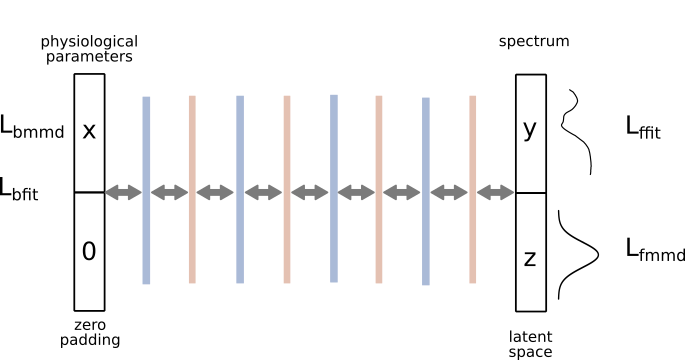}
    \caption{Schematic view of the network architecture applied in this paper. Blue layers correspond to invertible layers and orange layers correspond to permutation layers. The $L_{\ast}$ denote the loss functions used.}
    \label{fig:net}
\end{figure}

\paragraph{Basic principle} INNs have been proposed recently as a new method to recover a posterior distribution $p(\mathbf{x} | \mathbf{y})$ from an observation $\mathbf{y}$ \cite{ardizzone_analyzing_2018}. 
The network takes the form of a deterministic function $g$:
$$
\mathbf{x} = g(\mathbf{y}, \mathbf{z}; \Theta) \text{ with } \mathbf{z} \sim p(\mathbf{z}) = \mathcal{N}(\mathbf{z}; 0, I),
$$
where $\Theta$ denotes the trainable parameters and $\mathbf{z}$ are the \emph{latent variables} carrying the uncertainty of the reconstruction of $\mathbf{x}$ given $\mathbf{y}$. Sampling the latent variables according to the normal distribution  $\mathcal{N}(\mathbf{z}; 0, I)$ yields an approximation of $p(\mathbf{x} | \mathbf{y})$. 

\paragraph{Application to physiological parameter estimation} In the context of physiological parameter estimation, we hypothesize that observing a spectrum $\mathbf{y}$ is generally not sufficient to recover the underlying tissue parameter(s) $\mathbf{x}$. Intuitively speaking, the purpose of the latent variables \textbf{$\mathbf{z}$} is to capture the information necessary to recover $\mathbf{x}$ that is not already captured by $\mathbf{y}$. To recover a physiological parameter from a previously unseen spectrum $\mathbf{y}^\text{test}$, we repeatedly draw samples $(\mathbf{z}_i)_i$ from the latent space to obtain samples ($\mathbf{y}^\text{test},\mathbf{z}_i$) that we pass through the network. The corresponding set of physiological parameters $\mathbf{x}_i$ yields the posterior $p(\mathbf{x} | \mathbf{y}^{test})$. Due to the invertible architecture, the network simultaneously learns (1) the forward model - i.\,e., how to convert tissue parameters to spectral reflectances as measured by a camera -  and (2) how to recover a posterior distribution $p(\mathbf{x} | \mathbf{y})$ of tissue parameters corresponding to an observation $\mathbf{y}$.

\paragraph{Network architecture} The network architecture applied in this work has been adapted from~\cite{ardizzone_analyzing_2018} and can be found in Figure~\ref{fig:net}. It relies on four invertible affine coupling blocks~\cite{dinh2016density}, each of which is followed by a permutation layer, leading to an eight layer network in total. The purpose of the permutation layer is to improve the mixing of the different input and output channels. It adds no additional weights to the network. At initialization, a randomly chosen permutation between the input and output channels is fixed permanently. We assume that each physiological parameter has its own uncertainty associated. Hence, we choose $\dim z = \dim x = 13$ in this study.

\paragraph{Loss functions} Four loss terms are used to train the network (cf. Figure~\ref{fig:net}):
\begin{description}
\item[$L_2$ forward loss ($L_\text{ffit}$):] In the forward direction, we use an $L_2$ loss $L_\text{ffit}$ on the predicted reflectances $\mathbf{\hat y}$ and the true reflectances $\mathbf{y}$ to enforce good estimations for the forward process.
\item[MMD forward loss ($L_\text{fmmd}$):] We apply a Maximum Mean Discrepancy (MMD) loss $L_\text{fmmd}$ on the latent space estimation $\mathbf{\hat z}$. MMD losses distinguish between distributions~\cite{gretton2012kernel}. Here, we compare the distribution of the predicted latent variables $\mathbf{\hat z}$ to latent variables $\mathbf{z}$ sampled from the desired distribution $\mathcal{N}(\mathbf{z}; 0, I)$. 
\item[$L_2$ backward loss ($L_\text{bfit}$):] We use the estimations $\mathbf{\hat y}$ and $\mathbf{\hat z}$ from the forward pass and perturb both quantities with additive Gaussian noise. The resulting output $\mathbf{\hat x}$ with zero padding $\hat 0$ is compared to $\mathbf{x}$ with zero padding $0$ via the $L_2$ loss $L_\text{bfit}$.
This serves as a form of regularization, smoothing the latent space and ensuring that no critical information is hidden in low-amplitude structures in the outputs.
\item[MMD backward loss ($L_\text{bmmd}$):] We compute a reverse pass through the net with reflectances $\mathbf{y}$ from the training set $T^\text{train}$ and latent variables $\mathbf{z}$ from $\mathcal{N}(\mathbf{z}; 0, I)$. The output $\mathbf{\hat x}$ is then passed to an MMD loss $L_\text{bmmd}$, which compares it to the distribution given by the training samples $\mathbf{x}$. As previous work~\cite{ardizzone_analyzing_2018} indicates that $sO_2$, $v_\text{HB}$ and $a_\text{mie}$ are the only tissue parameters that can potentially be recovered from multispectral measurements, we decided to only feed these slices of $\mathbf{\hat x}$ into $L_\text{bmmd}$ instead of the whole prediction. 
\end{description}

\paragraph{Hyperparameter optimization} We use the training data set $T^\text{train}$ to perform the parameter optimization of the network and the validation data set $T^\text{validation}$ to prevent overfitting and for hyperparameter tuning. Particularly, we use the validation data to \emph{calibrate} the width of the posterior distributions.
As suggested in~\cite{niculescu2005predicting}, the purpose of the calibration is that for every sample, the $\alpha$-confidence interval ($\SI{0}{\%}\leq \alpha \leq \SI{100}{\%}$) of the posterior contains the ground truth value in $\alpha$ of the cases. In other words, for each value of $\alpha$ exactly a fraction $\alpha$ of the ground truth values shall be \emph{inliers} of the corresponding $\alpha$-confidence interval. We optimize the parameters using the validation set $T^\text{valid}$ to enforce this behavior as best as possible.


\subsection{Performance assessment}
\label{ssec:modes}

We quantify the uncertainty of an inference based on two key parameters: The presence of \textit{multiple} modes and the width of the posterior.

 Given samples following the posterior $p(\mathbf{x} | \mathbf{y})$ our approach to automatic mode detection relies on computing a kernel density estimation for $p(\mathbf{x} | \mathbf{y})$ which has the advantage of being easily sampled. This then allows us to compute the corresponding relative maxima of $p(\mathbf{x} | \mathbf{y})$. A posterior is classified as \emph{multi-modal}, if its standard deviation is less than half of the prior's standard deviation, our algorithm finds more than one relative maximum, and these maxima are further than a certain threshold apart ($sO_2$: \SI{3}{pp}, $v_\text{HB}$: \SI{0.3}{pp}). Furthermore, maxima whose intensity is less than 80\% of the main (i.\,e.\ highest) maximum are ignored. All remaining posteriors are classified as uni-modal.

To assess the performance of a camera, the INN is applied to $T^\text{test}$, and the automatic mode detection (Section~\ref{ssec:modes}) is run on each posterior $p(\mathbf{x} | \mathbf{y}^\text{test})$. Next, the following metrics are computed:
\begin{itemize}
    \item Percentage of multiple modes (MM): The percentage of multi-modal posterior distributions. We do this as a means to judge how well-posed the inversion is for the different cameras.
    \item Root-mean-square error (RMSE): We utilize maximum a posteriori probability (MAP) estimate as a predictor for the physiological parameters and give the root-mean-square error of these estimations against the ground truth.
    \item 68\% confidence interval width (W): We report the median interval width of the 68\% confidence interval as a measure of the width of the posterior distributions.
\end{itemize}

\section{Experiments and Results}
The purpose of the experiments was to confirm the realism of our simulation pipeline (Section~\ref{ssec:realism}) and to apply our setup to the task of comparative camera assessment (Section~\ref{ssec:comparative}).

\subsection{Realism of Simulation Pipeline}
\label{ssec:realism}
The simulation pipeline applied for comparative camera assessment features two potential sources of error: (1) Errors in the conversion of the simulated high resolution spectrum to multispectral measurements (sec.~\ref{sssec:camera}) and (2) wrong model assumptions in the generic tissue model and hence errors in the simulated spectra (sec.~\ref{sssec:tissue}). We will address both issues in this order in the following paragraphs.

\subsubsection{Virtual Camera}
\label{sssec:camera}
\begin{figure}[htbp]
    \centering
    \includegraphics[width=\textwidth]{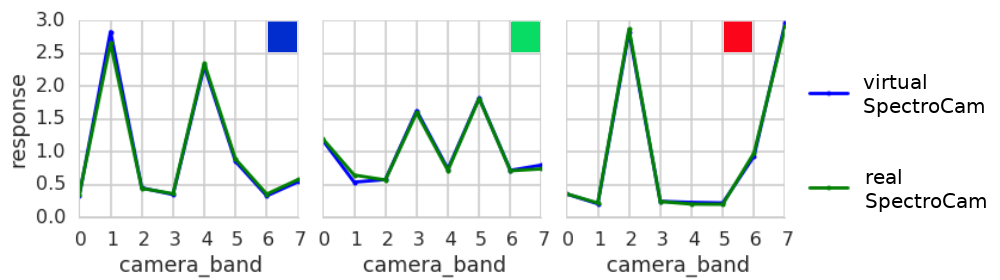}
    \caption{From left to right measurements of a blue, green and red color tile. The green line depicts the measurement by the actual camera (Pixelteq SpectroCam™). The blue line depicts the measurements of the virtual SpectroCam™ generated from the spectrometer measurements.}
    \label{fig:color_tiles}
\end{figure}

The realism of the simulated data relies crucially on the validity of our virtual cameras. In order to explore this, we measured color tiles (X-Rite ColorChecker®classic, Grand Rapids, MI, USA) which have a well defined spectrum using a HR2000+ spectrometer (Ocean Optics, Largo, FL, USA) and a Pixelteq SpectroCam™, which is a 8 band multispectral camera. Using the filter response functions of the SpectroCam, we transformed the high resolution spectrum to a virtual SpectroCam spectrum. To perform this experiment we used three color tiles (blue, green and red) and averaged five SpectroCam measurements. The measured intensities were normalized. As shown in Figure~\ref{fig:color_tiles}, the simulated data is in very close agreement with the real measurements.

\subsubsection{Tissue Model}
\label{sssec:tissue}
\begin{figure}[htbp]
    \centering
     \begin{subfigure}[b]{0.48\textwidth}
        \centering
        \includegraphics[width=\textwidth]{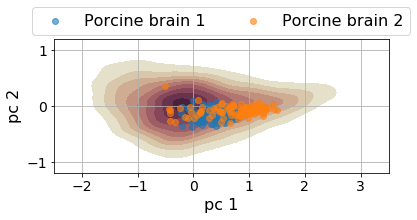}
       \caption{ }
       \label{fig:pig-pca}
    \end{subfigure}
     \begin{subfigure}[b]{0.48\textwidth}
        \centering
        \includegraphics[width=\textwidth]{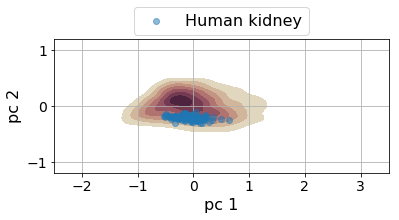}
       \caption{ }
       \label{fig:human-pca}
    \end{subfigure}
    \caption{Projection of measured organ data on the first two principal components (pc) of the simulated data (contour plot). (a) Measurements of two porcine brains with an 8-band camera. (b) Measurement of a human kidney with a 16-band camera.}
    \label{fig:organs}
\end{figure}

Due to the lack of a reliable gold standard method for measuring optical tissue properties \emph{in vivo}, validation of the tissue model is not straightforward. Previous work has addressed this issue by comparing real measurements of tissue with simulated spectra~\cite{wirkert_physiological_2017}. If the accuracy of the \textit{virtual camera} used can be assumed to be acceptable, deviations between real and simulated data can primarily be attributed to differences in the tissue composition. The tissue model applied in this study has been  validated in a previous publication \cite{wirkert_physiological_2017} using multispectral data from several different porcine abdominal organs. To confirm these findings we additionally acquired measurements from a porcine brain and a human kidney. The brain was measured using the same SpectroCam as in Section~\ref{sssec:camera}. For the human kidney, we used a 16-band camera. We performed a principal component analysis (PCA) on the data generated by our tissue model (adapted to the appropriate camera) and a kernel density estimation (KDE) on the first two principal components. Afterwards, we projected the measured data on those same components. The result can be found in Figure~\ref{fig:organs}. Clearly, all the organ data points lie within the distribution of the simulated data of our tissue model.

\subsection{Comparative Camera Assessment}
\label{ssec:comparative}
The main purpose of our experiments was to evaluate our assessment framework in the specific context of multispectral camera selection for physiological parameters estimation.

\subsubsection{Experimental Data} 
\label{ssec:cameras}

\begin{figure}
    \centering
    \includegraphics[width=\textwidth]{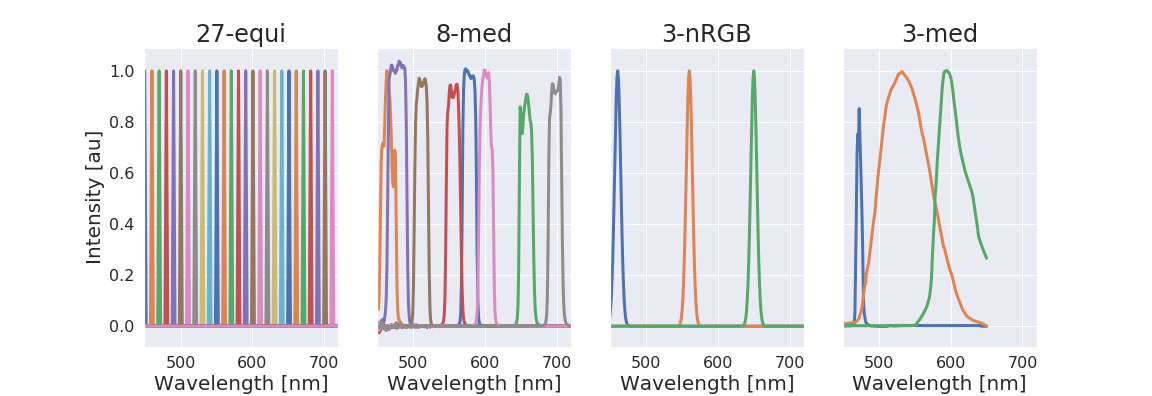}
    \caption{The filter response functions $f_i$ for the four cameras considered in this study.}
    \label{fig:filter_response}
\end{figure}

 We applied the Monte Carlo-based method described in Section~\ref{ssec:data} to generate 20,000 data points representing spectral reflectances using the tissue parameter distributions summarized in Table~\ref{tab:sim_colon}. If we give a range for a parameter this parameter is sampled uniformly from this range. If we give a value with standard deviation this parameter is sampled according to a normal distribution with this expectation value and standard deviation. For each camera setup investigated here, these reflectances were converted into (simulated) camera measurements considering the optical properties of the setup. For each setup, we reserved 70\% of the data for training ($T^\text{train}$), 5\% for hyperparameter tuning ($T^\text{validation}$) and 25\% for performance assessment ($T^\text{test}$).
 To test our assessment framework, we assessed three camera designs that have been applied in previous work~\cite{kaneko2014hypoxia,moccia2018uncertainty,wirkert_robust_2016} in a comparative manner. To obtain a lower bound on the achievable uncertainty, we complemented these realistic cameras by a virtual camera with nearly optimal design. The cameras are characterized by the following filter responses $f_i$ (cf.\  Figure~\ref{fig:filter_response}):
 
\begin{table}[htbp]
        \centering
        \begin{tabular}{c c c c c c c c}
            \toprule
            layer & $v_{\textrm{HB}} \lbrack \% \rbrack$           & $sO_2 \lbrack \% \rbrack$  & $a_{\text{mie}} \lbrack \si{\per\centi\meter} \rbrack$ & $b$         & $g$      & $n$       & $d \lbrack\si{\milli\meter}\rbrack$\\ \midrule
            1 & 0-10 & 0-100 & $18.9\pm10.2$ & 1.286 & 0.8-0.95 & 1.33 & 0.06-0.1\phantom{00}\\
            2 & 0-10 & 0-100 & $18.9\pm10.2$ & 1.286 & 0.8-0.95 & 1.36 & 0.06-0.085\\
            3 & 0-10 & 0-100 & $18.9\pm10.2$ & 1.286 & 0.8-0.95 & 1.38 & 0.04-0.06\phantom{0}\\
            \midrule
            \multicolumn{8}{l}{$\mu_a(v_{\text{HB}}, s, \lambda) = v_{\text{HB}} (s \epsilon_{\text{HBO2}}(\lambda) + (1-s)\epsilon_{\text{HB}}(\lambda)) \ln (10)(150\si{\gram\per\liter})(64,500\si{\gram\per\mole})^{-1}$}\\
            \multicolumn{8}{l}{$\mu_s(a_{\text{mie}}, b, \lambda) = \frac{a_{\text{mie}}}{1-g}(\frac{\lambda}{500\si{\nano\meter}})^{-b}$}\\ \midrule
            \multicolumn{8}{l}{framework: MCML\cite{alerstam2010next}, $10^6$ photons per simulation}\\
            \multicolumn{8}{l}{wavelength range $(\Lambda)$: 450-\SI{720}{\nano\meter} (stepsize=\SI{2}{\nano\meter)}}
        \end{tabular}
        \caption{The physiological parameter ranges used for simulating the desired tissue model.}
        \label{tab:sim_colon}
\end{table}


 

\begin{description}
\item[\emph{3-med}:] 3-band camera optimized for medical imaging use, as described in~\cite{kaneko2014hypoxia}.
\item[\emph{3-nRGB}:] 3-band camera whose bands' centers coincide with the standard RGB bands, as described in~\cite{moccia2018uncertainty}.
\item[\emph{8-med}:] Pixelteq SpectroCam, a 8-band-camera optimized for medical imaging use, as described in~\cite{wirkert_robust_2016}. This is the same camera as used in the experiments in Section~\ref{ssec:realism}.
\item[\emph{27-equi}:] As a close to optimal camera, we used a camera with a filter response featuring a (unrealistically) narrow band every \SI{10}{\nano\meter}. As our experimental data is based on presimulated data in the range of \SI{450}{\nano\meter} to \SI{720}{\nano\meter}, this leads to a `27 band camera'.
\end{description}

The remaining parameters $l(\lambda)$ and $o(\lambda)$ were set to 1 for all cameras.

\subsubsection{Results}
\label{ssec:results}

\begin{figure}[htbp]
    \centering
    \begin{subfigure}[b]{0.3\textwidth}
    \centering
    \includegraphics[width=\textwidth]{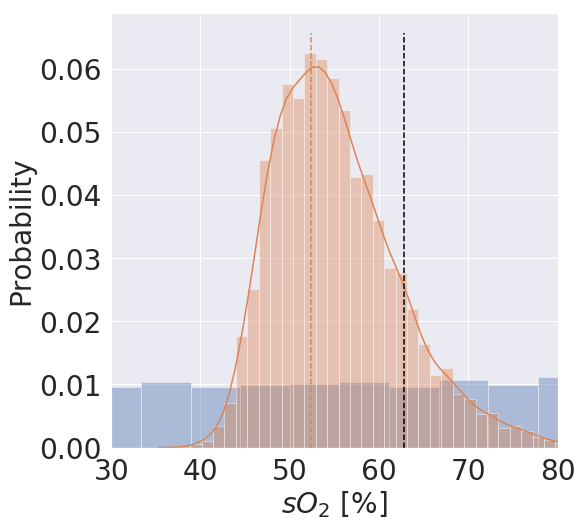}
    \caption{}
    \end{subfigure}
    \begin{subfigure}[b]{0.3\textwidth}
    \centering
    \includegraphics[width=\textwidth]{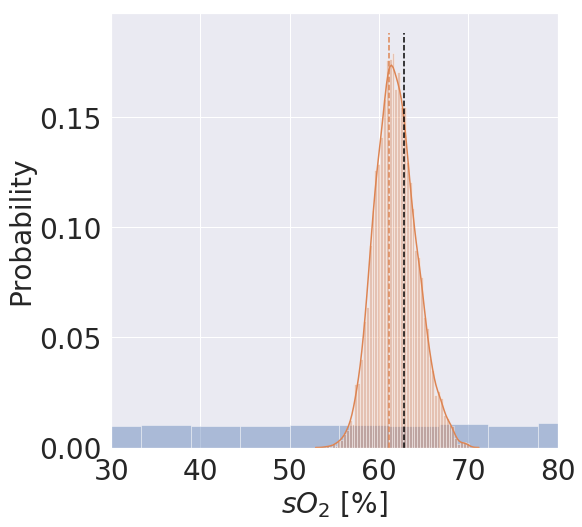}
    \caption{}
    \end{subfigure}
    \begin{subfigure}[b]{0.3\textwidth}
    \centering
    \includegraphics[width=\textwidth]{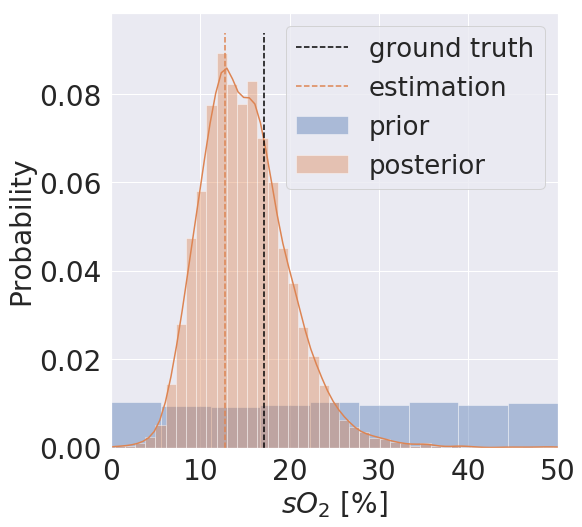}
    \caption{}
    \end{subfigure}
    \caption{Examples of INN output. (a) Uni-modal posterior with high width of the 68\% symmetric confidence interval (W), (b) uni-modal posterior with low W, (c) multi-modal posterior with two modes in close proximity.}
    \label{fig:posteriors}
\end{figure}
\begin{figure}
    \centering
    \includegraphics[width=0.8\textwidth]{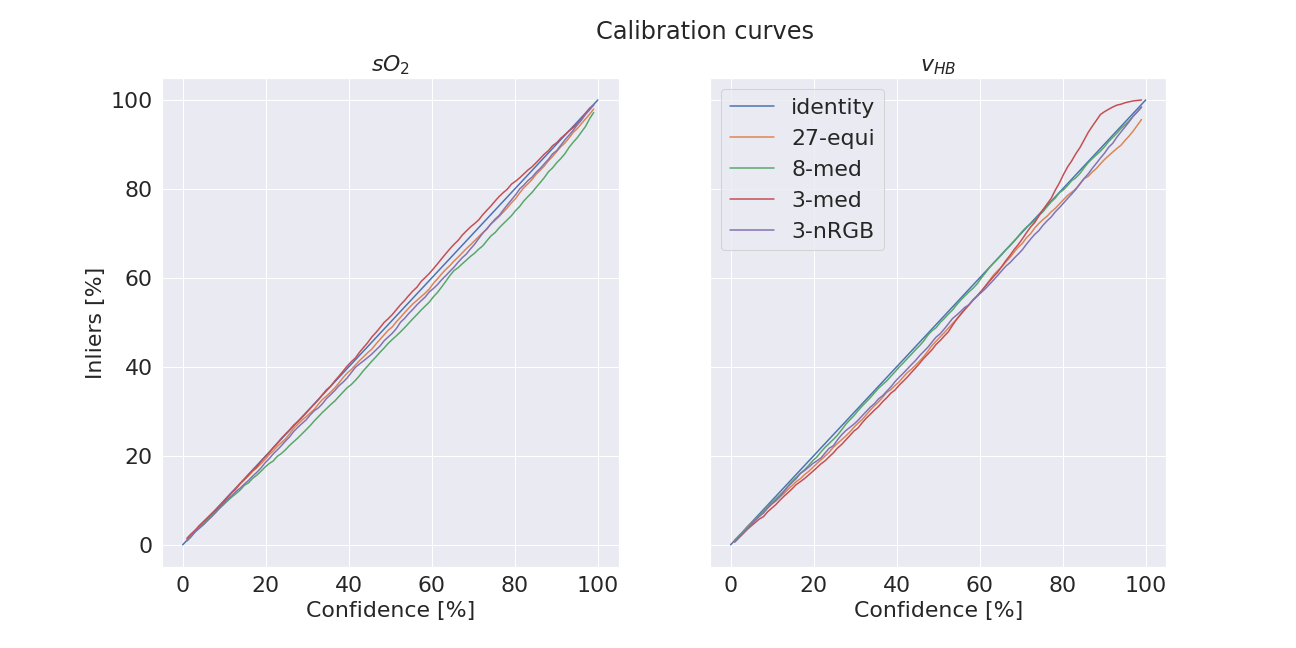}
    \caption{Calibration curves for tissue oxygenation ($sO_2$; left) and blood volume fraction ($v_{\text{HB}}$; right)}
    \label{fig:calibration}
\end{figure}

 Figure~\ref{fig:posteriors} provides representative examples for the posteriors generated by our INN. The calibration errors for the four different cameras are presented in Figure~\ref{fig:calibration} for the physiological parameters tissue oxygenation ($sO_2$) and blood volume fraction ($v_{\text{HB}}$). 
We see that the calibration curves closely follow the identity. In the case of $v_\text{HB}$, the 3-med camera is `underconfident' for larger values which would make estimations based on the confidence intervals in this range less reliable.


\begin{table}[htbp]
    \centering
    \sisetup{
        table-format = 1.1
    }
    \begin{tabular}{r c c c c c c}
        \toprule
         Camera &  \multicolumn{3}{c}{$sO_2$} & \multicolumn{3}{c}{$v_{\text{HB}}$}\\
         & {MM [\%]} & {RMSE [pp]} & {W [pp]} & {MM [\%]} & {RMSE [pp]} & {W [pp]}\\
         \midrule
        27-equi & 0.3 & 2.3  & 2.3 (0\%) & 0.1 & 1.6 & 3.4 (67\%)\\
        8-med & 0.5 & 2.9  & 4.0 (0\%) & 0.0 & 1.7  &  3.5 (62\%)\\ 
        3-nRGB & 9.3 & 4.8 & 5.8 (0\%) & 0.0 & 1.7 & 3.1 (64\%) \\ 
        3-med & 3.6 & 5.7 & 8.6 (0\%) & 0.0 & 2.4 & 5.3 (99\%)\\ 
        \bottomrule
    \end{tabular}
    \caption{Results for the comparative camera assessment. MM: Percentage of multi-modal posteriors; RMSE: Root-mean-square error; W: Median width of the 68\% symmetric confidence interval; pp: percentage points; The value in brackets denotes the percentage of samples with $W \ge 0.5 \cdot \sigma_{prior}$, where $\sigma_{prior}$ represents the standard deviation of corresponding the prior distribution. Note that a prerequisite for a distribution to be classified as multi-modal is  $W < 0.5 \cdot \sigma_{prior}$.}
    \label{tab:rms}
\end{table}

Table~\ref{tab:rms} shows the performance of the four different cameras using the  metrics presented in Section~\ref{ssec:modes}. All computations were performed on the test set $T^\text{test}$. As expected, the scores generally improve with an increasing number of spectral bands.  An interesting observation is that the 3-band camera designed for medical use (\emph{3-med}) has a higher RMSE compared to the camera whose design was inspired by standard RGB cameras (\emph{3-nRGB}), yet, it features a substantially  reduced number of multiple mode posteriors (3.6\% vs 9.3\%).

\begin{figure}[htbp]
    \centering
    \begin{subfigure}[b]{0.48\textwidth}
        \centering
        \includegraphics[width=\textwidth]{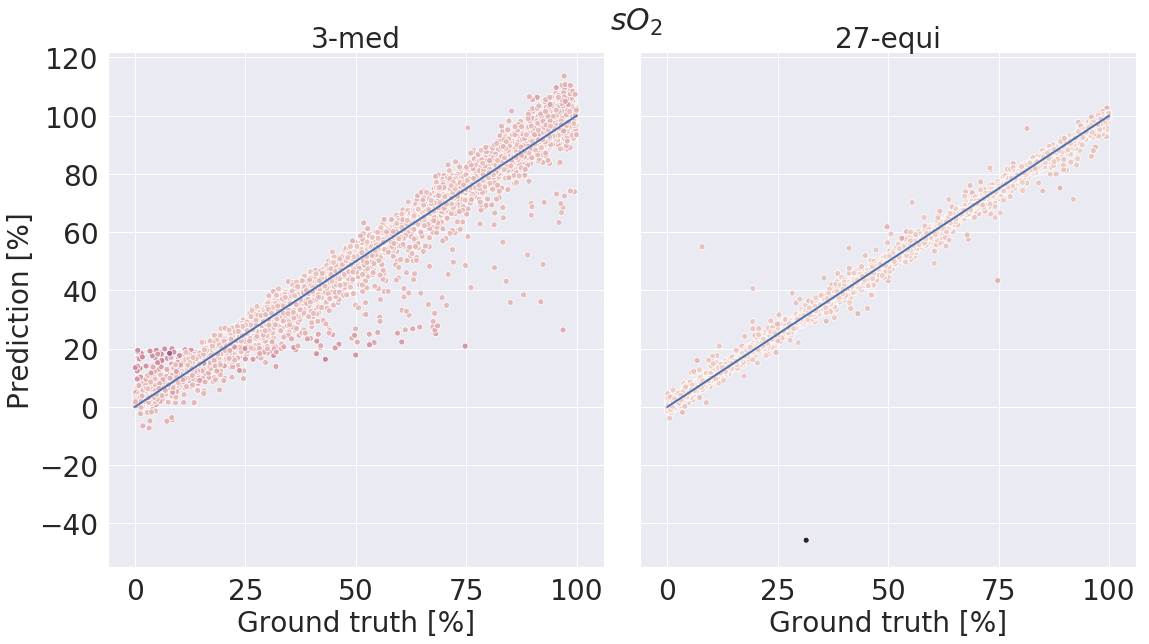}
        \caption{}
        \label{fig:so2}
    \end{subfigure}
    \begin{subfigure}[b]{0.48\textwidth}
        \centering
        \includegraphics[width=\textwidth]{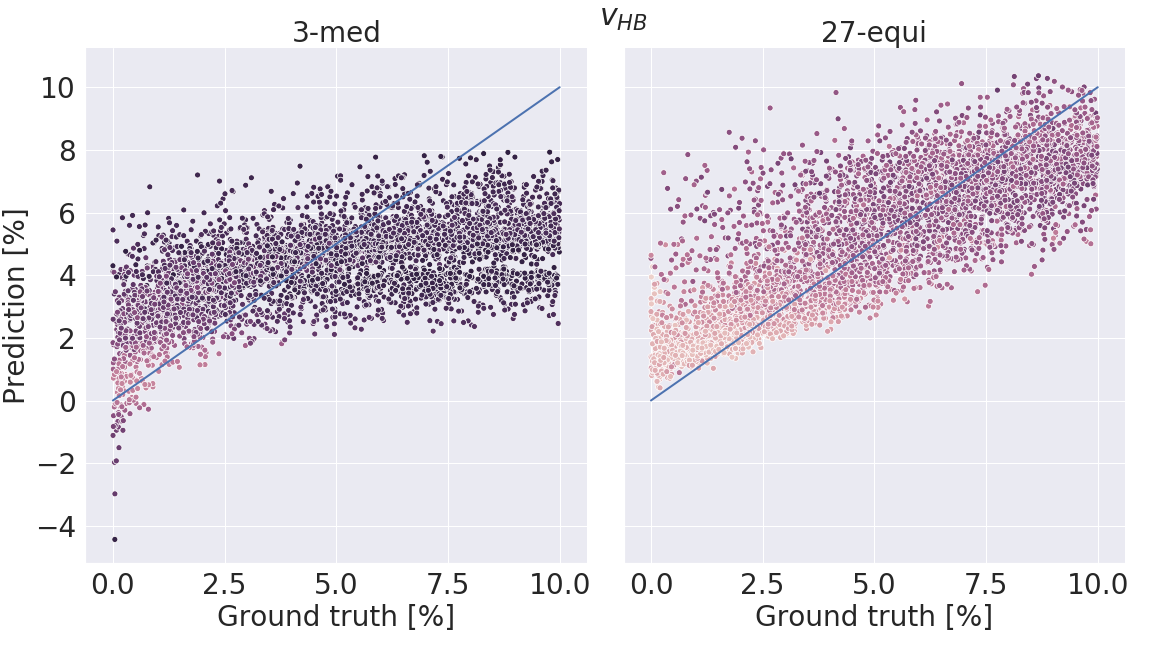}
        \caption{}
        \label{fig:vhb}
    \end{subfigure}
    \caption{Worst (3-med; left) and best (27-equi; right) cases for $sO_2$ and $v_\text{HB}$ estimation according to the RMSE. The hue represents the width of the 68\% confidence interval (c.\,f.\ Table~\ref{tab:rms}). We see that both cameras have difficulties to predict higher $v_\text{HB}$ values, but encode this in the width of the posterior. The blue line depicts the identity mapping.}
\end{figure}

For all cameras except the \textit{3-nRGB}, there are only few multiple mode posteriors for $sO_2$ reconstruction. Figure~\ref{fig:so2} shows the estimations of the \textit{27-equi} and the \textit{3-med} camera which show generally good reconstructive performance and the possibility of outlier detection via the width of the posteriors.


In contrast, our results suggest that  $v_\text{HB}$ cannot be recovered from any of the cameras with high certainty. In fact, the percentage of samples with $W \ge 0.5 \cdot \sigma_{prior}$, (where $\sigma_{prior}$ represents the standard deviation of the corresponding the prior) is greater than 50\% for all four cameras. The poor performance is illustrated in Figure~\ref{fig:vhb}. We see that the \emph{27-equi} camera still performs better than the \emph{3-med} camera, but
none of them show good performance for high $v_\text{HB}$ values. This general trend is also true for the other two cameras. Note that since most posteriors were even wider than the priors, they did not qualify as a candidate for the multiple mode mode detection algorithm (cf.\ Section~\ref{ssec:modes}) explaining the low MM.

Furthermore, although W seems reasonable in absolute terms, comparing it to twice the standard deviation of the prior distribution reveals that the median width of the posteriors goes as high as 93\% for the \emph{3-med} camera, indicating that $v_\text{HB}$ is effectively unrecoverable. For $sO_2$ the values range from 4\% in the \emph{27-equi} case to 15\% in the \emph{4-med} case. 

\section{Discussion}
Meaningful performance assessment and benchmarking are crucial for advancing research and practice. Several publications, however (cf.\ e.\,g.~\cite{maier-hein2018}), suggest that the metrics chosen are not always well-suited for a specific assessment goal. In the context of multispectral intra-operative imaging, for example, camera assessment has typically been restricted to determining descriptive statistics on error metrics that quantify the difference between the estimations of an algorithm and reference (\textit{gold standard}) results~\cite{waibel2018,wirkert_robust_2016,wirkert_physiological_2017}. An advantage of this approach is that the error metrics are straightforward to compute and interpret. On the other hand, such performance measures suffer from the fact that they do not reveal important insights with respect to why methods perform poorly. In particular, they do not account for the different types of uncertainties that may occur when recovering tissue parameters from camera measurements. An interesting practical example is the 3-band camera designed for medical use~\cite{kaneko2014hypoxia} and investigated here. While it features a higher RMSE compared to a 3-band camera based on the standard RGB design, recovery of tissue parameters is substantially less ambiguous, as indicated by the reduced number of multiple modes. 

To address the issues related to commonly applied approaches to camera design, selection and performance assessment, we present a novel approach to camera assessment which provides the following key advantages compared to previously proposed methods:
\begin{enumerate}
    \item Extended scope: The topic of camera design is closely linked to that of band selection \cite{wirkert18_domain_confer_presen}. To our knowledge, however, none of the approaches proposed in this field addresses the potential inherent ambiguity associated with the recovery of physiological parameters. To overcome this bottleneck, we propose moving beyond point estimates and mapping measurements to a full posterior probability distribution. Analysis of the posteriors not only provides us with a means for quantifying the uncertainty related to a specific measurement but also allows for a fundamental theoretical analysis about which tissue properties can in principle be recovered with the present camera.
    \item No need for acquisition of real data: Many approaches to band/camera selection rely on acquisition of real data \cite{gu_image_2016,han_vivo_2016,nouri_efficient_2014,nouri_hyperspectral_2016,wirkert_endoscopic_2014}. Yet, acquisition of real data for a given application is often impractical due to budget constraints (no money to purchase a whole range of cameras) or ethical issues. We address this issue by performing the comparative assessment \textit{in silico}. Experiments with a whole range of porcine and human organs confirm the realism of our simulation framework. 
\end{enumerate}

The above computations show that these networks can compute the same error metrics as before (e.\,g.\ RMSE) while having the potential for finer differentiation through additional metrics (e.\,g.\ number of modes or width of posterior). 

While it is straightforward to compute the widths of the posteriors, fully-automatic multiple mode detection is not trivial due to the many parameters involved. For example, the posteriors are only implicitly given by a number of samples generated according to a latent space sample. This fact alone introduces statistical fluctuations into the estimated posterior. A kernel density estimation can smoothen out these effects, but at the cost of introducing a bandwidth parameter with a high impact on the number of resolved maxima. In addition, outliers must be handled in order to avoid faulty signals at the boundary of the posterior. 

The calibration of our models suggest that while the confidence of our posteriors is already good, there is still room for improvement. In particular, the calibration of $v_\text{HB}$ for the 3-med camera is off for larger confidences. In future studies which aim at finer differentiation between the observed cameras this would have to be remedied. We are confident that this can be achieved keeping in mind the convincing results for the other three cameras.

Another obstacle which learned methods have to sidestep are the so called \emph{out of distribution samples}. The performance of our algorithm can only be guaranteed on data that is similar to the training data. In general, this problem is difficult to tackle. In our case, the PCA projections of the organ measurements show exemplary that the spectra of many interesting objects, like internal organs, are in fact in our training distribution.

The color tile experiments together with the measured organ spectra suggest the validity of our simulation framework. A natural next step would be to test the performance of our method on real data. To achieve this, there remains a key challenge: real data will always be subject to noise which needs to be handled adequately by our algorithm. One approach would be to average the spectra either by using a higher integration time or by averaging multiple measurements. However, if there are time constraints, for example induced by organ movement, there are limits to the amount of averaging possible. Another approach would be to incorporate a realistic noise model in the simulation framework to account for it during training. This would circumvent the time constraint as the evaluation time of the network trained with this new data set would not change.

Another interesting direction for future work is to apply the framework to additional cameras that are widely used in a clinical context (e.\,g.\ RGB or narrow band cameras). We expect some obstacles with regard to the extension to 2-band cameras as there is just very little information left for a multi parameter reconstruction. Additionally, these cameras would need a larger range of simulated wavelengths compared to the data set that we based our work on~\cite{wirkert_robust_2016}. However, extending the framework to these ranges should be straight forward.

Additionally, while this study focused on intra-operative optical imaging, the concept of performance assessment using INNs could easily be transferred to other fields of research. Clearly, any imaging modality with pixelwise spectral information is a prime candidate. For larger image context, the INNs are still in active development. Because of their peculiar structure, the hidden layer size is the same as the input and output dimension leading to very large networks when images are to be processed as a whole. One example of an imaging modality where it might be fruitful to apply our INN method is the field of quantitative photoacoustic imaging (qPAI). It has been shown before that qPAI is an ill-posed inverse problem in theory~\cite{cox2009challenges}. However, to the best of our knowledge an \emph{in silico} or even \emph{in vivo} analysis of the practical implications of this non-uniqueness has not been conducted. The ability to detect ambiguous reconstructions of physiological parameters seems like a promising candidate to close this gap.

In conclusion, we have presented a novel method for performance assessment of optical cameras  bearing the potential to measure the well-posedness of the inverse problem. Future work should focus on the evaluation steps necessary to fully harness the power of the computed posterior distributions. In particular, robust mode detection algorithms seem like a fruitful area for further investigation in order to quantify the uniqueness of the reconstruction.


\renewcommand{\abstractname}{\ackname}
\begin{abstract}
This study has received funding from the European Union’s Horizon 2020 research and innovation programme through the ERC starting grant {\footnotesize COMBIOSCOPY} under grant agreement No ERC-2015-StG-37960.

Conflict of Interest: The authors declare that they have no conflict of interest.

All procedures performed in studies involving human participants were in
accordance with the ethical standards of the institutional and/or national
research committee and with the 1964 Helsinki declaration and its later
amendments or comparable ethical standards. All applicable international,
national, and/or institutional guidelines for the care and use of animals were
followed. All procedures performed in studies involving animals were in
accordance with the ethical standards of the institution or practice at which
the studies were conducted. For this type of study formal consent is not
required.

\end{abstract}


%
%

\end{document}